\documentclass[preprint,review,12pt]{elsarticle}
\usepackage{graphicx}
\usepackage{epsf}
\usepackage{amsmath}
\usepackage{lscape}
\usepackage{color}
\usepackage{dcolumn}
\usepackage{hyperref}
\usepackage{amsmath}
\usepackage{rotating}
\usepackage{soul}
\usepackage{longtable}

\newcommand{\cm}{cm$^{-1}$}
\newcommand{\p}{$^\prime$}
\newcommand{\pp}{$^{\prime\prime}$}

\newcommand{\NH}{NH$_{3}$}

\newcommand{\N}{N$_{2}$}

\newcommand{\abinitio}{\textit{ab initio }}

\journal{Journal of Quantitative Spectroscopy \& Radiative Transfer}

\begin{document}

\begin{frontmatter}

\title{High-resolution absorption measurements of \NH\ at high temperatures: 2100 - 5500 \cm}
\author{Emma~J. Barton, Sergei.~N. Yurchenko, Jonathan Tennyson}
\address{Department of Physics and Astronomy, University College London,
London, WC1E 6BT, UK}
\author{S\o nnik Clausen, Alexander Fateev}
\address{Technical University of Denmark, Department of Chemical and Biochemical Engineering, Frederiksborgvej 399,
4000 Roskilde, Denmark}

\begin{abstract}

  High-resolution absorption spectra of \NH\ in the region 2100 - 5500
  \cm\ at 1027 $^{\circ}$C and  atmospheric pressure
  (1045 $\pm$ 3 mbar) are measured. An \NH\ concentration of 10\% in
  volume fraction is used in the measurements. Spectra are recorded in
  a high-temperature gas-flow cell using a Fourier Transform Infrared
  (FTIR) spectrometer at a nominal resolution of 0.09 \cm. The spectra
  are analysed by comparison to a variational line list, BYTe, and
  experimental energy levels determined using the MARVEL procedure.
  2308 lines have been assigned to 45 different bands, of which 1755
  and 15 have been assigned or observed for the first time in this
  work.

\end{abstract}

\end{frontmatter}

\section{Introduction}

\NH\ spectra can be used to extract physical information from
spectroscopic observations of a range of hot and cold environments. On
Earth \NH\ is an important component in several industrial process as
for example gasification and NOx reduction in combustion
\cite{11DTxxxx.NH3}.  Such processes can be monitored and optimised
with the help of \textit{in situ} measurement of gas temperature and
composition \cite{12FaClxx.industry}. In space \NH\ is ubiquitous and
used to probe, for example, circumstellar envelopes
\cite{16ScSzBu.NH3}, star-forming regions \cite{16LaKiTs.NH3}, dense
molecular clouds \cite{16HaDaSi.NH3}, the atmospheres of cool stars
\cite{jt596}, brown dwarfs \cite{15LeMoCa.NH3} and giant solar system
planets \cite{77WoTrOw.NH3}. Recent work includes the first detection
of gas-phase ammonia in a planet-forming disk \cite{16SaHoBe.NH3}.

Many experimental studies have focused on the \NH\ molecule providing,
for example, high temperature experimental line lists
\cite{12HaLiBea,12HaLiBeb}, ro-vibrational assignments
\cite{jt508,jt616,jt633,14CeHoVeCa} and experimentally derived
energies \cite{12HaLiBea,12HaLiBeb,jt508,jt616,jt633,14CaCeCoRo}. A
comprehensive compilation of measured \NH\ rotational and
ro-vibrational spectra can be found in a recent MARVEL study
\cite{jt608}. The MARVEL (measured active rotation-vibration energy
levels) algorithm \cite{jt412,12FuCsi.method} simultaneously analyses
all available assigned and labelled experimental lines, thus yielding
the associated energy levels.  Al-Derzi analysed
29~450 measured  \NH\ transitions and yielded 4961 accurately-determined
energy levels which mostly lie below 7000 \cm\ \cite{jt608}.  Very
recently Sung {\it et al} \cite{16SuYuPePi} have significantly
improved the spectral coverage for ammonia in the far infrared (50 -- 660 \cm).

The broad temperature and spectral range of applications can be
difficult to cover exhaustively in the lab because of NH$_3$ thermal
decomposition either in the gas phase or on the walls of a gas cell
(heterophase).  To help fill in the gaps a number of theoretical line
lists have been computed for \NH\ \cite{jt466,jt500,11HuScLe2.NH3}. In
the present work a variationally computed line list for hot \NH, BYTe
\cite{jt500}, is employed. This line list covers the spectral range 0
- 12,000 \cm\ and is expected to be fairly accurate for all
temperatures up to 1500 K (1226 $^{\circ}$C). In particular, BYTe
shows errors in band origins which can be up to a 3 or 4 \cm\ for 
bands involving high-lying vibrational states \cite{jt633,jtKP2} but
can be expected to be lower for the region studied here and
to extrapolate smoothly with $J$ for a given band.
BYTe comprises of
1~138~323~251 transitions constructed from 1~373~897 energy levels
lying below 18~000 \cm. It was computed using the NH3-2010 potential
energy surface \cite{jt503}, the TROVE ro-vibrational computer program
\cite{07YuThJe.method} and an \abinitio dipole moment surface
\cite{jt466}. However a new line list currently being constructed as
part of the ExoMol project \cite{jt528,jt631} as BYTe is known to have
some problems reproducing experimental intensities \cite{jt616,jt633}
and is less accurate for higher wavenumber transitions
\cite{jt633,11HuScLe,11HuScLe2,12SuBrHuSc}. Assigned high resolution
laboratory spectra are needed to refine and validate theoretical line
positions and intensities.

In our previous study \cite{jt616} we extended work by Zobov {\it et
  al.}~\cite{jt508} by analysing new hot absorption spectra in the
region 500 - 2100 \cm. In the current work we present and analyse new
hot absorption spectra in the region 2100 - 5500 \cm.  It should be
noted that high temperature (up to 1400 $^{\circ}$C) experimental line
lists for the region 2100 - 4000 \cm\ are available due to Hargreaves
et al.  \cite{12HaLiBeb} based on their observed emission spectra.

This article has the following structure. The experimental set-up
used for the measurements is described in Section 2. Section 3 gives
an overview the assignment procedure and the method used to calculate
experimental and theoretical absorbance spectra. The accuracy of BYTe
is assessed by a direct comparison with the experimental
spectra in Section 4.1 and summary of all assignments is given in
Section 4.2. Finally our conclusions are presented in Section 5.

\section{Experimental Details}

The experimental setup is described in our previous work
\cite{jt616}, the main points are summarised below.

\begin{figure}[htbp]
\centering
\includegraphics[width=12cm]{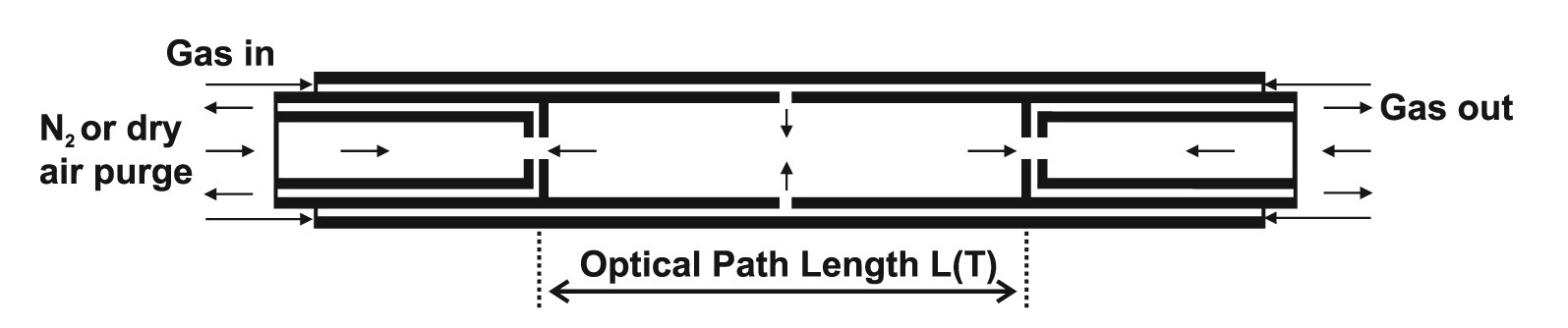}
\caption{Schematic of the gas flow patterns in the high temperature ceramic 
gas cell (c-HGC) used in the measurements. 
The optical path length is defined by flow windows formed 
by counterwise gas flows (N$_2$ and gas of interest). See [34] for more details.  Black arrows indicate flow direction.  Reproduced from Ref.
\cite{15AlWeMa.NH3}.}
	\label{f:gascellc}
\end{figure}

\begin{figure}[htbp]
	\centering
	\includegraphics[width=13cm]{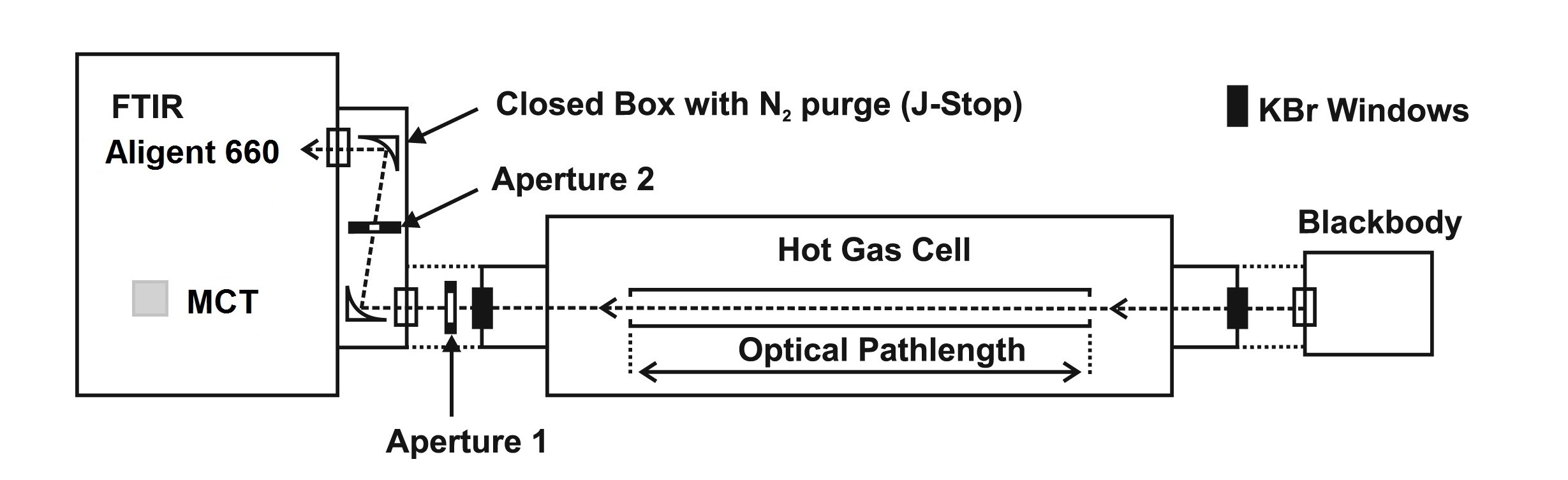}
	\caption{Experimental optical set up for NH$_3$ high-resolution measurements  in the c-HGC. . Adapted from Ref.~\cite{15AlWeMa.NH3}.}
	\label{f:opticalsetup}
\end{figure}

An Agilent 660 FTIR spectrometer, linearised Mercury-Cadmium
Telluride (MCT) detector, ceramic high-temperature gas-flow cell
(c-HGC) (see Figure~\ref{f:gascellc}) and an external IR light source,
which is Blackbody-like (BB) at 1800 K were used in the measurements.
The optical setup is illustrated in Figure~\ref{f:opticalsetup}.

The c-HGC operates at temperatures up to 1873 K (1600 $^{\circ}$C)
\cite{15ClNiFa.industry} and has also been used by the Technical
University of Denmark (DTU) group
\cite{12EvFaCl.NH3,11BeClFa.industry,15AlWeMa.NH3} to study for
example hot CO, CO$_2$, CH$_4$ and H$_2$O. This cell has a
fully-heated, temperature-uniform central part and two
partially-heated buffer parts with interchangeable optical (KBr)
windows at the ends. The buffer parts are purged with \N\ or dry air
taken from a purge generator while the sample gas (e.g. \N\ + \NH) is
preheated and fed into the central part of the cell. Laminar flow
sheets (flow windows) are established between the central and buffer
parts where the purge and sample gases meet, meaning the sample gas
can not reach, or react with or form deposits on, the optical windows
\cite{08FaClxx.industry}.  To minimise reactions with the internal
surface of the gas cell the inner part of the c-HGC is made from high
quality pure ceramic (Al$_2$O$_3$(99.5$\%$)). The absorption path
length, defined by the flow windows, has a value of 53.3 cm at room
temperature. At higher temperatures the length changes a little due to
thermal expansion. Thus at 1027 $^{\circ}$C the length is 53.8 cm that is 
an increase of about
0.9\%\ of its value at room temperature, see
Ref.~\cite{11BeClFa.industry}.

An NH$_3$ (9.94\%) concentration in N$_2$ (99.998\%) was obtained by mixing
N$_2$ (99.998\%) and NH$_3$ (99.98\%) flows controlled using a
high-end (BRONKHORST) mass-flow controllers (MFC). The
N$_2$ (99.998\%) bottle was obtained from Air Liquide and the NH$_3$
(99.98\%) bottle was obtained from Linde Gas.  The remaining gas in the
NH$_3$ (99.98\%) bottle was considered to be air.  The accuracy of the MFC
was taken to be $\pm 0.9$\%\ of the reading plus $\pm 0.2$\%\ of the full
scale. Therefore for the N$_2$ the MFC accuracy under measurement
conditions was $\pm 2.4$\%. If other than calibration gas is used, one
needs to add the uncertainty in the conversion factor from calibration gas
to actual gas which is in our case is 3.6\% (for NH$_3$ instead of Ar).
Therefore the Ar MFC (used with NH$_3$) has an accuracy of 6.2%.

Further details on the c-HGC, its performance and a comparison with
the other HGCs in the laboratory will be presented elsewhere
\cite{15ClNiFa.industry}.  For now the reader is referred to
Ref.~\cite{15AlWeMa.NH3}.

Single beam (SB) spectra from measured interferograms at a nominal
spectral resolution 0.09 cm$^{-1}$ are calculated using Agilent
Resolution Pro software (supplied with the FTIR spectrometer) using
inverse fast Fourier transform (FFT) and boxcar and triangular
apodization functions. Mertz phase correction is applied, see
Griffiths and de Haseth's book \cite{07Grdexx.book}. Triangular
apodization results in less noise in the final spectra while boxcar
apodization gives narrower peaks.  The boxcar apodization allows one
to maximize the possible FTIR spectrometer performance in the sense of
spectral resolution. If the true line widths are at least 5 times
greater than the spectral resolution of the FTIR, then the measured
line profiles can be considered as true ones and therefore more
fundamental studies about line shape can be carried out. This is the
case, for example, for high-pressure measurements with a FTIR of
similar class to ours. This is also the case when spectra are broad
(i.e. with a continuum-like structure).  For our spectra (and for PNNL
in general) there is no big difference between the boxcar and
triangular apodizations, except for very closely-spaced lines, because
one needs to model spectrum at the FTIR resolution in any case, even
for 0.112 \cm\ resolution.  To ensure consistent results both sets of
calculated SB spectra were used in the final analysis.

Measured wavenumbers were multiplied by a factor of 1.000059 to
account for the linear wavenumber shift caused by beam divergence, in
accordance with the discussion in Ref.~\cite{15AlWeMa.NH3}.  The
experimental uncertainties on absorbance measurements were determined
by comparison of two high-resolution NH$_3$ spectra measured on two
separate days and estimated to be 2.9~\%\ for 3050 -- 3650 \cm,
5.1~\%\ for 4200 -- 4600 \cm\ and 7.7~\%\ for 4860-4900 \cm.  An
effective value for the experimental uncertainty in the absorbance
measurements for the whole spectrum of 5.2~\%\ was adopted.

\section{Data Analysis}

This study used the BYTe \cite{jt500} variational line list
and experimental energies determined using the MARVEL procedure \cite{jt608}.

\subsection{Calculating Experimental Absorption Spectra}

Experimental transmission spectra $\tau_{\rm exp}(\nu,T)$ at a
temperature $T$ [K] and a line position $\nu$ [\cm] are calculated
from four SB spectra.  This four measurements scheme is needed because
the IR light source used is not modulated but modulation of the IR light
appears in the FTIR spectrometer. Therefore, the MCT detector in the FTIR
spectrometer will "see" both modulated emissions from the light source
and the cell. Moreover at 1027~$^{\circ}$C many of NH$_3$ bands appear in emission
as well and the additional measurements with the cold beam stopper allow one
to separate emission from absorption. The four measurements are
two reference (\N\ in the central part of the cell) measurements $\mathbf{I}_{\rm ref+BB}$
and $\mathbf{I}_{\rm ref}$ and two sample (\N\ + \NH\ mixture) measurements
$\mathbf{I}_{\rm gas+BB}$ and $\mathbf{I}_{\rm gas}$, one with
and one without signal from the BB (at 1800 K) \cite{jt616,12EvFaCl.NH3}:

\begin{equation}
\tau_{\rm exp}(\nu,T) = \frac{\mathbf{I}_{\rm gas+BB} - \mathbf{I}_{\rm gas}}{\mathbf{I}_{\rm ref+BB} - \mathbf{I}_{\rm ref}}
\end{equation}

\noindent
Spectra without signal from the BB are measured
from a cold (room temperature) beam stopper placed at 90 degrees from the optical axis of the setup using a movable mirror in the BB adapter.
The absorption spectra are then calculated from the reference,
$a_0$ ($=\mathbf{I}_{\rm ref+BB} - \mathbf{I}_{\rm ref}$), and
sample, $a_1$ ($=\mathbf{I}_{\rm gas+BB} - \mathbf{I}_{\rm gas}$), measurements:

\begin{equation}
A_{\rm exp}(\nu, T) = \log_{10} \left[\frac{a_0}{a_1}\right]
\end{equation}

\subsection{Calculating Theoretical Absorption Spectra}

The method for calculating theoretical absorption spectra follows Ref.~\cite{jt616}.
First the 'true' transmission spectrum was computed as:

\begin{equation}
\tau^{\rm true}_{\rm calc}(\nu, T) = \exp\left(-\sigma(\nu, T) l c \right)
\end{equation}

\noindent
where $l$ is the absorption path length in cm, $c$ is the \NH\ concentration in cm$^{-3}$
and $\sigma(\nu, T)$ is the pressure-broadened \NH\ absorption cross-section calculated using BYTe
and the procedure laid out in \cite{jt542}, but replacing the Gaussian line shape with a Voigt line shape.
Lorentz half-widths were estimated from the experimental
spectra and with reference to measured widths compiled in the HITRAN database.
The measured (effective) transmittance spectrum is derived by convolving $\tau^{\rm true}_{\rm calc}(\nu, T)$
with the instrument line shape (ILS) function $\Gamma(\nu - \nu_0)$:

\begin{equation}
\tau^{\rm eff}_{\rm calc}(\nu, T) = \int_{0}^{\infty} \tau^{\rm true}_{\rm calc}(\nu_0, T) \Gamma(\nu - \nu_0) d\nu_0
\end{equation}

\noindent
For boxcar apodization, the ILS is a $\textrm{sinc}$ function:

\begin{equation}
\Gamma(\nu) = \Lambda~{\rm sinc}(\Lambda\pi\nu) = \Lambda\frac{\sin(\Lambda\pi\nu)}{(\Lambda\pi\nu)}
\end{equation}

\noindent
For triangular apodization, the ILS is a $\textrm{sinc}^2$ function:

\begin{equation}
\Gamma(\nu) = \Lambda~{\rm sinc}^2(\Lambda\pi\nu) = 2\Lambda\frac{\sin^2(\Lambda\pi\nu)}{(\Lambda\pi\nu)^2}
\end{equation}

\noindent
where $\Lambda$ is commonly termed the FTIR retardation and is generally defined as the inverse of
the nominal resolution of the spectrometer \cite{07Grdexx.book}.

\noindent
The theoretical absorption spectrum is then computed as:

\begin{equation}
A_{\rm calc}(\nu, T) = \log_{10} \left[\frac{1}{\mathbf{\tau}^{\rm eff}_{\rm calc}(\nu, T)}\right]
\end{equation}

\subsection{The Assignment Procedure}

First a list of observable BYTe lines for the experimental conditions
was compiled. For this purpose the absorbance of each line, $j$, was
approximated as:

\begin{equation}
	A^{\rm approx}_{\rm calc} = \frac{S_{j}^{a} l c}{\Delta L \ln(10)}
%\log_{10} \left[\frac{1}{\exp\left(-S_{j}^{a} l c \right)}\right]
\end{equation}

\noindent
where $\Delta L$  is an effective line width which is assumed to be a constant for all lines in the spectrum and the quantity
 $\frac{S_{j}^{a}}{\Delta L}$  represents an effective cross section assuming rectangular line shapes with
$\Delta L$ widths.
%is the absorption coefficient in cm/molecule \cite{jt338}. Lines with $A^{approx}_{calc} > $ 0.001 were kept.

If both the upper and lower energies involved in a observable
transition were known experimentally, the BYTe line position was
replaced by the MARVEL line position generated by subtracting upper
and lower state energies. This hybrid line list, which retains
all the BYTe transitions,
will be presented elsewhere \cite{jtNH3hot}; it shall henceforth be
referred to as BARVEL. 

Taking the resolution of the measurements and the accuracy of BYTe intensities
into account (see Section 4.1), experimental peaks and BARVEL line positions were coupled
using python scripts to produce a 'trivial' assignment list. In cases
where multiple BARVEL lines corresponded to a single peak, the peak was
assigned to the strongest line.

Trivial assignments for the same vibrational band provide an expected
observed minus calculated (obs. - calc.) difference for all lines in that band.
Lines present in the list of observable BYTe lines, but not in BARVEL, were shifted by this
residual to make future assignments by the method of branches \cite{jt205}.

A list of all trivial and branch assignments,
the final assignment list, was then compared to previous studies,
namely those catalogued in the HITRAN database \cite{jt557}.

\section{Results and Discussion}

The absorption measurements were performed at a temperature of
1027 $^{\circ}$C for the \NH\ volume concentration of 10$\%$.

The measurements were used to test the accuracy of BYTe then analysed
using BYTe to generate an assignment list for the data. Central wavenumbers
for assigned peaks are compared to line positions measured by Hargreaves et al. \cite{12HaLiBeb} where possible.

The absorption spectra, a peak list (partially assigned) including line
positions from Hargreaves et al. \cite{12HaLiBeb} for assigned lines  where available, and new energy level information
derived from the assignments are presented in the supplementary data.

\subsection{Direct Comparison with BYTe}

\begin{figure}[htbp]
\centering
\scalebox{0.5}{\includegraphics{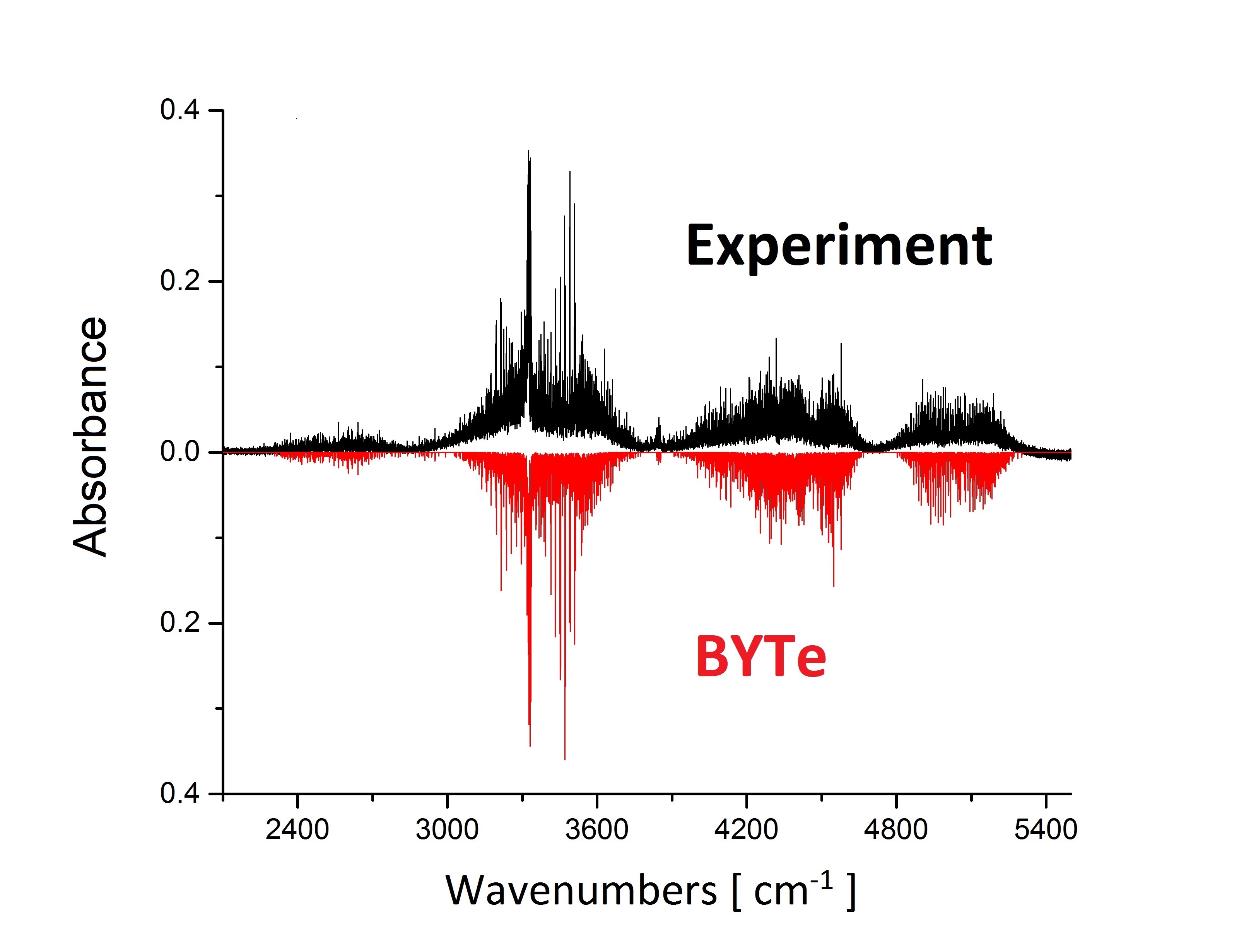}}
\caption{Comparison between experimental (upper) and calculated (BYTe, lower) absorption spectra at 1027 $^{\circ}$C
	for the range 2100 - 5500 \cm.  }
\label{f:byte1}
\end{figure}

\begin{figure}[htbp]
\centering
\scalebox{0.5}{\includegraphics{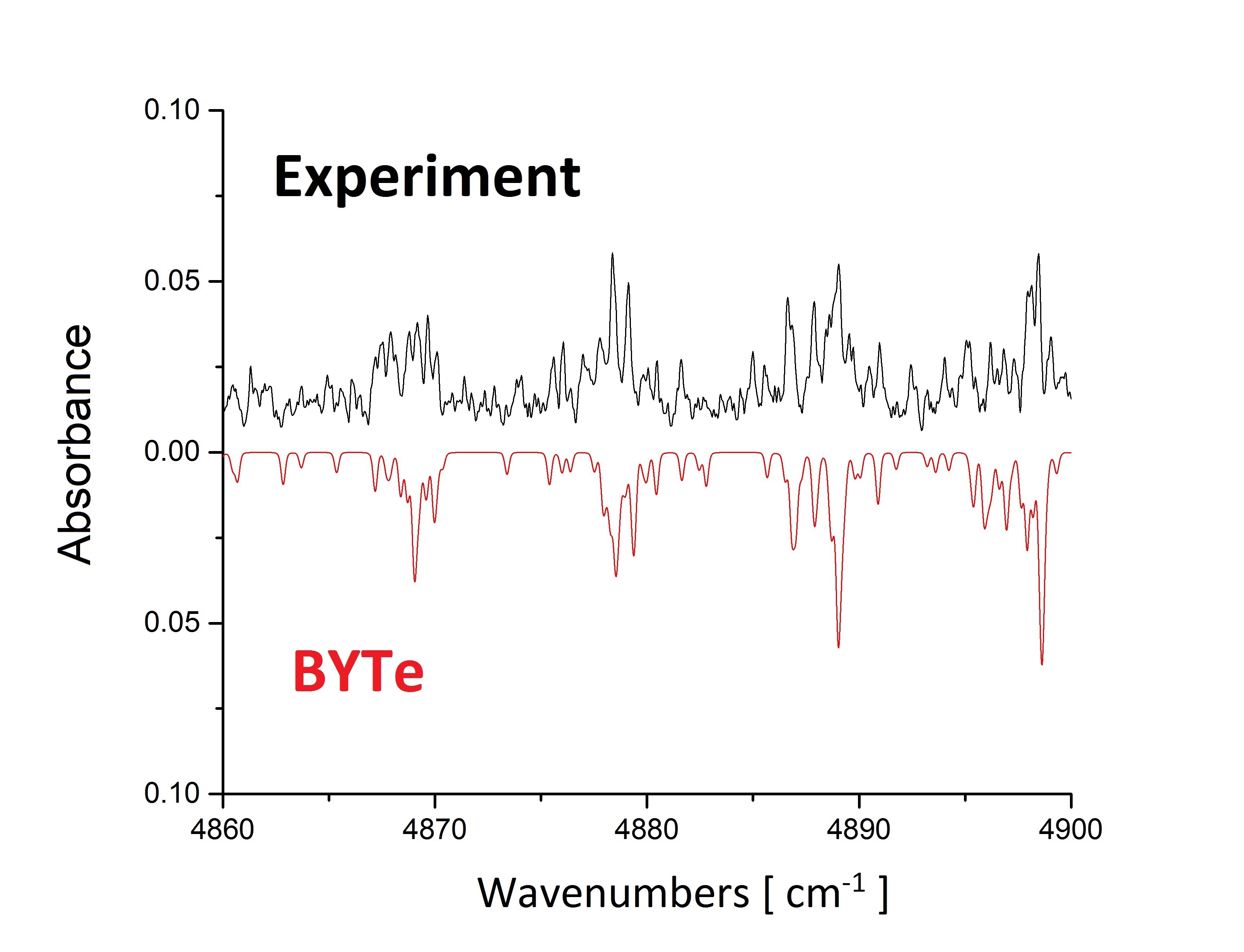}}
\caption{Comparison between experimental (upper) and calculated (BYTe, lower) absorption spectra at 1027 $^{\circ}$C
	for the range 4860 - 4900 \cm.}
\label{f:byte2}
\end{figure}

A comparison between the experimental and theoretical absorption spectra at 1027 $^{\circ}$C
for the whole region (2100 - 5500 \cm) is shown in Figure~\ref{f:byte1}. Overall, taking
into account the experimental noise, there is good agreement. However there are shifts in
line position of the order $\pm$ 0.2 \cm\ across the entire spectral range and shifts up to
$\pm$ 1 - 2 \cm\ in a few regions, particularly at higher wavenumbers. Hence
it was decided that assignments should only be made using MARVEL line positions or BYTe line
positions corrected for the expected obs. - calc. difference derived from trivial assignments,
and not by simple line list comparison. BARVEL line positions should have an obs. - calc. difference
smaller than the nominal resolution of the measurements, 0.09 \cm, whilst the wavenumber threshold
for the BYTe line positions was taken to within 0.1 \cm\ of the expected obs. - calc. difference.
On the whole experimental line intensities are reproduced within 30 \%. This is illustrated in
Figure~\ref{f:byte2} for the region 4860 - 4900 \cm. As such experimental lines were coupled to
BARVEL or BYTe lines using an intensity threshold of 30 \%.

\subsection{Assignments}

Out of 3701 measured experimental peaks 2308 lines
have been assigned. The remaining peaks either
did not correspond to a BARVEL or BYTe line within the set
wavenumber and intensity thresholds or corresponded
to multiple lines with roughly equal contribution
to the total intensity such that it could not be
confidently assigned. 553 lines were previously assigned
by studies included in the HITRAN database (see Table~\ref{t:assignments2}).
The full 1027 $^{\circ}$C
peak list with assignments is available as supplementary
material to this article.

Hargreaves et al. \cite{12HaLiBeb} presented high temperature line lists for the
region 2100 - 4000 \cm\ constructed from emission spectra recorded
at a resolution of 0.01 \cm. These line lists are currently being updated
and extended (private communication)
and hence were not the focus of the current work. Of the 1755 newly assigned
lines in this work, 990 are also present in the line lists of Ref.~\cite{12HaLiBeb}.
In these cases line positions from Ref.~\cite{12HaLiBeb} are included with the current
central peak wavenumbers in the supplementary data and employed in the computation of
upper state energies described below, as these were measured at a higher resolution.

For branch assignments with an experimentally known lower energy
state, energies for the upper state were computed using MARVEL
energies and the line position of the strongest assigned transition
to that state. The calculated energies are available as supplementary
material to this article.

\begin{table}[htbp]
\caption{Summary of \NH\ lines assigned in the region 2100 - 5500 cm$^{-1}$.}
\begin{center}
\begin{tabular}{lr}
\hline
&  Lines     \\
\hline
Experimental   &  3701 \\
HITRAN  &   553   \\
New trivial  & 272 \\
New branch   &  1483 \\
Total Assigned & 2308 \\
\hline
\end{tabular} \label{t:assignments2}
\end{center}
\end{table}

As in our previous study \cite{jt616},
lines were assigned to a large number of different bands.
Table~\ref{t:bands} gives a summary of the observed bands
including the number of lines assigned to each and whether the
band was observed for the first time in this work. Bands are listed
in order of theoretical vibrational band centre (VBC).
 VBC = VBO\p - VBO\pp\ where VBO is the vibrational band origin from BYTe,
in wavenumbers. For simplicity abbreviated vibrational labels
($v_1 v_2 v_3^{L_3} v_4^{L_4} i$) \cite{jt546} are used to identify
bands in this table and only the highest value of the
rotational quantum number $J$, assigned in this work for
each band, is indicated. If the observed $J_{\mathrm{max}}$ in
this work is bigger that quoted in the literature, the previous
$J_{\mathrm{max}}$ is also given. The full 26 quantum labels for
each transition, 13 per vibration-rotation state as
recommended by Down {\it et al.}~\cite{jt546},
will be given in the partially assigned peak list and energies files.

\begin{center}
\begin{longtable}{llrcrcl}
\caption[Summary of observed NH$_3$ bands: 2100 - 5500 \cm.]{Summary of observed bands
in the region 2100 - 5500 \cm\ in order of theoretical (BYTe) vibrational
band centre (VBC = VBO\p\ - VBO\pp\ where VBO = vibrational band origin, in \cm)
with  maximum upper and lower $J$ rotational quantum number
($J^\prime_{\mathrm{max}}$ and $J^{\prime\prime}_{\mathrm{max}}$ respectively).
$N$ is the number of lines assigned to the band.
If $J_{\mathrm{max}}$ in this work is higher than given in the literature, the
previously known $J_{\mathrm{max}}$ is given in parentheses. 
o$-$c gives the band shift used, in \cm, in making branch assignments.
VBO of $0^+$ is set to 0.0 \cm\ in line with
the MARVEL study \cite{jt608}.} \label{t:bands} \\

\hline \multicolumn{1}{c}{Band} & \multicolumn{1}{c}{VBC} & \multicolumn{1}{c}{$N$} & \multicolumn{1}{c}{$J^\prime_{\mathrm{max}}$} & \multicolumn{1}{c}{$J^{\prime\prime}_{\mathrm{max}}$} & \multicolumn{1}{c}{o$-$c}& Note \\ \hline
\endfirsthead

\multicolumn{6}{c}%
{{\bfseries \tablename\ \thetable{} -- continued from previous page}} \\
\hline \multicolumn{1}{c}{Band} &
\multicolumn{1}{c}{VBC} &
\multicolumn{1}{c}{$N$} &
\multicolumn{1}{c}{$J^\prime_{\mathrm{max}}$} &
\multicolumn{1}{c}{$J^{\prime\prime}_{\mathrm{max}}$} &
\multicolumn{1}{c}{o$-$c}&
\multicolumn{1}{c}{Note}  \\ \hline
\endhead

\hline \multicolumn{6}{r}{{Continued on next page}} \\ \hline
\endfoot

\hline \hline
\endlastfoot
$\nu_3^{1,-} - \nu_2^-$                   & 2475.50& 59& 17 (12)&    16 &$-$0.1 \\
$\nu_3^{1,+} - \nu_2^+$                   & 2511.55& 52& 19 (12)   & 18 &0.1 \\
$(\nu_2 + \nu_3^1)^- - 2\nu_2^-$            & 2553.27& 14& 16 (11) & 15 &0.1 & New \\
$(\nu_2 + \nu_3^1)^+ - 2\nu_2^+$            & 2553.27& 6& 9  & 9        &0.1 & New \\
$3\nu_2^- - 0^+$                            & 2895.53& 6 & 9        & 9 &0.1 \\
$(\nu_1 + 2\nu_2)^+ - 2\nu_2^-$             & 3120.69& 8 & 16 (11)& 16  &$-$2.0 \\
$(\nu_2 + 2\nu_4^0)^+ - \nu_2^+$            & 3147.49& 1 & 8 & 8        &$-$0.6   & New \\
$(\nu_2 + 2\nu_4^2)^+ - \nu_2^+$            & 3167.81& 19 & 18 (7) & 19 &0.0   & New \\
$(2\nu_2 + \nu_4^1)^+ - 0^+$                & 3189.04& 17& 15 (11)  & 14&1.0   \\
$2\nu_4^{0,+} - 0^-$                        & 3215.21& 95& 21 (12)  & 20&0.3   \\
$2\nu_4^{0,+} - 0^+$                        & 3216.00& 2 & 6        &  7&0.3  \\
$2\nu_4^{0,-} - 0^-$                        & 3216.75& 13& 17 (12)  & 18&$-$0.2 \\
$2\nu_4^{0,-} - 0^+$                        & 3217.55&  72 & 18 (12)& 17&0.4 \\
$2\nu_4^{2,+} - 0^-$                        & 3239.39& 5 & 9        & 8 &$-$0.1   \\
$2\nu_4^{2,+} - 0^+$                        & 3240.18&  82 & 24 (13)& 25&$-$0.2   \\
$2\nu_4^{2,-} - 0^-$                        & 3240.78&  53 & 21 (13)& 22&$-$0.2   \\
$(\nu_2 + 2\nu_4^0)^- - \nu_2^-$            & 3240.81& 1 & 7 & 8        &0.0   & New \\
$2\nu_4^{2,-} - 0^+$                        & 3241.58&  23 & 17 (13)& 18&0.1   \\
$(\nu_2 + 2\nu_4^2)^- - \nu_2^-$            & 3260.70& 8 & 16 (8) & 17  &0.0   & New \\
$2\nu_1^+ - 2\nu_4^{0,-}$                   & 3296.58& 1 & 4      & 4   &1.0  & New \\
$(\nu_1 + \nu_2)^+ - \nu_2^-$               & 3326.39& 62& 22 (11)& 22  &0.05  \\
$(\nu_1 + \nu_4^1)^+ - \nu_4^{1,-}$         & 3328.35& 19& 14 (13)& 15  &0.0   & New \\
$(\nu_1 + \nu_4^1)^- - \nu_4^{1,-}$         & 3329.51& 11& 15 (12)& 16  &0.0  & New \\
$(\nu_1 + \nu_4^1)^- - \nu_4^{1,+}$         & 3330.61& 10& 13 (12)& 14  &0.0   & New \\		
$\nu_1^+ - 0^-$                             & 3335.28& 158 & 21 (12)& 21&$-$0.4   \\
$\nu_1^- - 0^+$                             & 3337.07& 155 & 21 (12)& 21&$-$0.4   \\
$(\nu_1 + \nu_2)^- - \nu_2^+$               & 3387.59& 85 & 19 (12)& 19 &$-$0.05   \\			
$\nu_3^{1,-} - 0^-$                         & 3443.20& 142 & 22 (12)& 22&0.1   \\	
$(\nu_3^1 + \nu_4^1)^+ - \nu_4^{1,+}$	    & 3443.60& 2 &   6& 7       &0.1  &New \\
$\nu_3^{1,+} - 0^+$                         & 3443.62& 160 & 23 (12) &24&0.1   & \\
$(\nu_2 + \nu_3^1)^+ - \nu_2^+$             & 3448.80& 110 & 19 (12)& 20&0.2   \\
$(\nu_2 + \nu_3^1)^- - \nu_2^+$             & 3467.32& 1& 7         & 7 &0.2  \\
$(\nu_2 + \nu_3^1)^- - \nu_2^-$             & 3503.01& 73 & 19 (11)& 20 &0.2   \\
$(2\nu_2 + \nu_3^1)^- - 2\nu_2^-$           & 3470.63& 18 & 14 (7)& 15  &0.3   & New \\
$(2\nu_2 + \nu_3^1)^+ - 2\nu_2^+$           & 3548.80& 74 & 21 (11)& 22 &0.0   & New\\
$(2\nu_2 + \nu_3^1)^+ - \nu_2^+$            & 4178.16& 89 & 18 (11)& 17 & $-$0.6        &  New\\
$(\nu_1 + \nu_2)^+ - 0^-$                   & 4293.73& 11 & 12 (11)& 12      & 0.05 \\
$(\nu_1 + \nu_2)^- - 0^+$                   & 4320.03& 12 & 13 (12)& 12     &0.0 \\
$(\nu_2 + \nu_3^1)^+ - 0^+$                 & 4416.93& 123& 19 (12)& 18  &0.2   \\
$(2\nu_2 + \nu_3^1)^- - \nu_2^-$            & 4420.38& 93 & 16 (7) & 17&0.4        &  New\\
$(\nu_2 + \nu_3^1)^+ - 0^-$                 & 4434.66& 129& 19 (11)& 18  &0.2 \\
$(\nu_1 + \nu_4^1)^+ - 0^+$                 & 4955.73& 100& 18 (13)& 18 &0.0 \\
$(\nu_1 + \nu_4^1)^- - 0^-$                 & 4956.10& 130& 19 (12)& 20 &$-$0.1 \\
$(\nu_3^1 + \nu_4^1)^- - 0^-$               & 5069.59& 3  & 9 (8)  & 9      &$-$1.0   \\
$(\nu_3^1 + \nu_4^1)^+ - 0^+$               & 5069.88& 1  & 8      &  9    &0.4 \\
\end{longtable}
\end{center}

15 bands have been assigned for the first time in this work, although some
of the energy levels involved are known from observations of other bands.

All trivial assignments are secure, as the MARVEL energies (and hence BARVEL
line positions) are known to very high accuracy (of the order 10$^{-4}$ \cm\ for the
energies). The accuracy of branch assignments depends on the determination of
the obs. - calc. difference for a given vibrational band.

For bands with many ($>$ 10) assignments the obs. - calc. difference can be tracked
through the band. As this remains relatively stable we have confidence in our assignments.

Bands for which only a few lines could be assigned are more tentative, although
every observed band in this work has at least one associated trivial assignment.

It is worth noting that the single lines assigned to $(\nu_2 + 2\nu_4^0)^+ - \nu_2^+$,
$(\nu_2 + 2\nu_4^0)^- - \nu_2^-$, $2\nu_1^+ - 2\nu_4^{0,-}$, $(\nu_2 + \nu_3^1)^- - \nu_2^+$
and $(\nu_3^1 + \nu_4^1)^+ - 0^+$ are all trivial.

\section{Summary}

High-resolution absorption measurements of \NH\ in the region 2100 - 5500 \cm\
at atmospheric pressure and a temperature of 1027 $^{\circ}$C have
been reported and analysed.

A comparison between the measurements and BYTe shows in general good
agreement through there are some shifts in line position (up
to 2 \cm) and overall BYTe reproduces experimental intensities only within
30 \%.  Work towards a new \NH\ line list is currently being
carried out as part of the ExoMol project \cite{jt528}.

The use of BYTe and MARVEL has allowed the assignment of 2308 lines. 553 lines
were previously assigned by studies included in the HITRAN database.
1755 lines have been assigned for the first time in this work.
The 272 lines assigned using MARVEL line positions, also known as trivial
assignments, are secure as the accuracy of MARVEL energies is of the order 10$^{-4}$ \cm.
Of the 1483 branch assignments, those associated with bands which have numerous assignments
in this work should be reliable because the observed-calculated differences remain
relatively stable within a given band. The remaining assignments should also be valid, as
all observed bands have at least one verified assignment in this work which provides an expected
observed-calculated difference for the band, however these are more tentative.

\section*{Acknowledgements}

This work was supported by a
grant from Energinet.dk project N. 2013-1-1027, by
UCL through the Impact Studentship Program and the European Research Council under Advanced Investigator Project 267219.

\bibliographystyle{elsarticle-num}

%\bibliography{journals_phys,industry,jtj,methods,NH3,NH3_MARVEL,linelists,CH4,Books}
%\bibliography{journals_phys,jtj,methods,NH3,NH3_MARVEL,linelists,CH4,book,industry}
%/data/tex/bib/jt

\end{document}